# Cascaded Förster Resonance Energy Transfer and Role of the Relay Dye

C. K. R. Namboodiri[*], P. B. Bisht[*], and V. R. Dantham[†]

*Department of Physics, Indian Institute of Technology Madras, Chennai, 600036, India.*

[*]*E-mail: bisht@iitm.ac.in*

*Tel. +91-44-2257-4866, Fax: +91-44-2257-4852*

The effect of introducing a relay dye on the energy transfer efficiency in a new donor-acceptor system has been studied. The values of the critical transfer distance, the reduced concentration, and the energy transfer efficiency of the cascaded system (consisting of three dyes- donor-relay dye-acceptor) are compared with that of the two dye system. Experimentally, it has been observed that the presence of a relay dye increases the transfer efficiency from the donor to the acceptor.

---

[**] Current Address: Department of Science& Mathematics, Regional Institute of Education Mysore, Mysore, 570006, India.
[†] Current Address: Department of Physics, Indian Institute of Technology Patna, Patna, 800013, India.

# 1. Introduction

Förster resonance energy transfer (FRET) is a distance dependent, non-radiative energy transfer based on the dipole-dipole interaction [1, 2]. FRET is being widely used in the study of protein-protein interaction [3, 4], protein folding [5-7], photosynthetic systems [8, 9], bioimaging [10, 11], and developing FRET based sensors [12-14]. Cascaded (multi-step or multi-color) FRET is a recent advance in this field and has advantages compared with the traditional two-color FRET such as wide range of absorption of sunlight [15] and its efficient conversion of light to the deep red region in luminescent solar concentrators (LSCs) [16], decreased photodegradation of the donor dyes [8], efficient single molecule detection [17], and simultaneous detection of multiple metal ions [18]. Here, light from the donor gets transferred to the acceptor mainly through a relay dye (Panel (B), Figure 1). Cascaded FRET is helpful for the conversion of light at a shorter wavelength to a much longer wavelength which has applications in solar cells. For such an efficient light conversion, the choice of a relay dye is crucial as it affects the overall transfer efficiency. Hence, the role of the relay dye on the energy transfer efficiency needs to be investigated. In view of this, steady-state as well as time-resolved measurements of a standard dye-pair along with a relay dye has been carried out. In this study, Calcein and DTTCI were chosen as the donor (D) and acceptor ($A_1$) dyes, respectively. DODCI ($R_1$) was chosen as the relay dye because of its absorption and fluorescence spectra falling in between the other two. For a comparison of the obtained results with this D-$R_1$-$A_1$ system, two more cascaded systems (Calcein- DODCI- HITCI and Calcein- Rhodamine B- HITCI) have been studied. This study gives experimental confirmation that the presence of a relay dye improves the energy transfer efficiency from the donor to the acceptor.

# 2. Experimental

Calcein (CALN), (BDH Ltd.), 3,3′-diethyloxadicorbocyanine iodide (DODCI), (Serva), 3,3′- diethylthiacarbocyanine iodide (DTTCI), (Serva), rhodamine B (RhB), (Serva), and 1, 1′, 3, 3, 3′, 3′- hexamethylindotricarbocyanine iodide (HITCI), (Sigma- Aldrich) were used as received. The molecular structure of the dyes used are reported. Spectrograde methanol (Thomas- Bakers) and distilled water (Modern Distilled Water, Chennai) were used as received. Polyvinyl alcohol (PVA, molecular weight ≈ 125000) was obtained from S. D. Fine-Chem. Ltd., Mumbai. 20 g of PVA in 200 ml of distilled water was stirred for half an hour at 300 K. Dye was mixed with PVA to obtain the uniform, bubble-free solution. PVA-dye solution was dispersed onto a slide glass, and was allowed to dry for 48 hours at room temperature to obtain the polymer samples. The entire FRET study was done using so obtained solid state PVA films of thickness around 100 μm.

Steady-state absorption spectra were recorded using a dual-beam absorption spectrometer (Jasco model V- 570). Fluorescence spectra were recorded using a Raman spectrometer (Horiba Jobin Yvon, HR 800) with a grating of 600 groves/mm. The samples were excited with 488 nm, $Ar^+$ laser. An objective lens of 100X was used for excitation. The fluorescence was collected with a Peltier cooled CCD (DU420A-OE-324, Andor Technologies) in a back scattering geometry after passing through an edge filter. The time of exposure was kept as 1 second for all measurements.

The time-resolved studies were done under *ps* excitation using the time-correlated single photon counting (TCSPC) technique, described elsewhere [19]. Briefly, a *ps* diode laser (50 *ps*, 470 nm, 1MHz, PiL047, Advanced photonics Systems, Germany) excites the sample and the fluorescence is collected in the back-scattering geometry by using an epifluorescence microscope (E400, Nikon) and is detected with a PMT (Hamamatsu R928). The decay curves were recorded with the help of a TCSPC card and software (T900, Edinburgh Instruments). Commercial software (FAST, Edinburgh Instruments) was used for data analysis. The goodness of the fit was judged by the value of the reduced chi-squared ($\chi^2$) and the distributions of residuals.

# 3. Theoretical Aspects

The rate of FRET is given by [20],

$$k_{D^*A} = \frac{1}{\tau_D}\left(\frac{R_{0A}}{r_{DA}}\right)^6 \quad (1)$$

where, $\tau_D$ is the fluorescence lifetime of the pure donor, and $R_{oA}$ is known as the critical transfer distance, and $r_{DA}$ is the real distance between the donor and acceptor molecules. Mostly, the fluorescence decay of an excited flourophore follows single exponential decay given by,

$$I_D(t) = I_0 \exp\left(-\frac{t}{\tau_D}\right) \quad (2)$$

In a donor-acceptor system, the donor decay follows the Förster expression given by [20],

$$I_D(t) = I_0 \exp\left[\left(-\frac{t}{\tau_D}\right) - 2\gamma_{DA}\left(\frac{t}{\tau_D}\right)^{\frac{1}{2}}\right] \quad (3)$$

where $\gamma_{DA}$ is the ratio of the acceptor concentration to the critical acceptor concentration ($\gamma_{DA} = \frac{C_A}{C_{0A}}$) known as the reduced concentration. Critical acceptor concentration is related to the critical transfer distance by,

$$C_{0A} = \frac{3000}{2\pi^{3/2} N R_{0A}^3} \quad (4)$$

The efficiency of the energy transfer ($\eta$) is calculated by [20],

$$\eta = \sqrt{\pi}\gamma_{DA} \exp(\gamma_{DA}^2)(1 - erf(\gamma_{DA})) \quad (5)$$

where the $erf(\gamma_{DA})$ is the Gaussian error function.

The decay of the acceptor follows the equation [20, 21],

$$I_A(t) = B_2 \exp\left(-\frac{t}{\tau_2}\right) - B_1 \exp\left(-\frac{t}{\tau_1}\right) \quad (6)$$

where $\tau_1$ and $\tau_2$ are the two lifetime components and $B_1$ and $B_2$ are the corresponding amplitudes.

## 4. Results & Discussions

### 4.1 Selection of dye system

Table I (a) gives the donor-donor (D-D) migration spectroscopic parameters of the dyes, used in this study. The dyes were so chosen that the overlap integral for D-D migration varies by three orders of magnitude, from $0.05 \times 10^{-13}$ M$^{-1}$cm$^3$ (for CALN) to $33.9 \times 10^{-13}$ M$^{-1}$cm$^3$ (for HITCI). The corresponding variation of the critical transfer distance ($R_{0self}$) was from $24 \pm 1$ Å to $69 \pm 1$ Å. The normalized absorption and normalized fluorescence spectra of the dyes in the polymer matrix (PVA) are shown in panel A of figure 1. It can be seen that the absorption and fluorescence spectra of these dyes cover the range of wavelengths from visible to near IR.

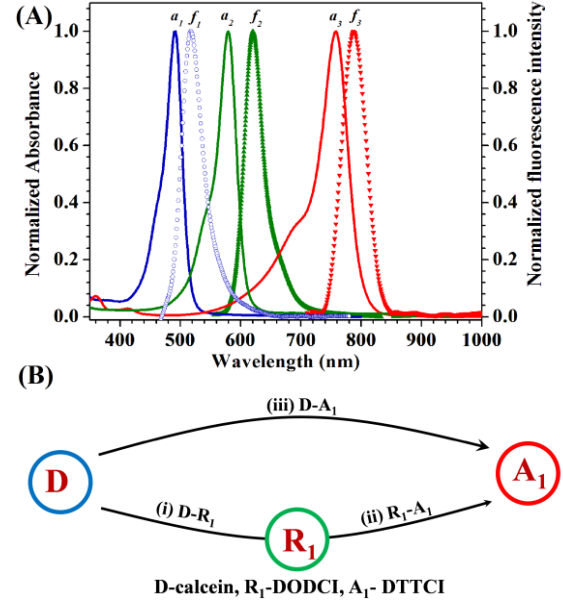

**Figure 1**. The normalized absorption (solid lines- $a_1$, $a_2$, and $a_3$) and the fluorescence (symbols- $f_1$, $f_2$, and $f_3$) spectra of CALN, DODCI, and HITCI respectively are shown in panel A. Panel B gives a schematic diagram of various FRET pathways in D-R$_1$-A$_1$ system.

Table I (b) gives the spectral overlap ($J_{DA}$) and $R_{0A}$ values for various dye-pairs. Besides the large value of $J_{DA}$ for CALN-RhB (D-R$_2$) and DODCI-DTTCI (R$_1$-A$_1$) systems, the values are small for other dye pairs because of wide separation between the fluorescence band of the donor and the absorption band of the acceptor.

### 4.2 Component dye pairs of the cascaded system

A cascaded system consists of 3 components, viz. D-R system, R-A system, and D-A system. Therefore, it is essential to study the properties of individual component systems

**Table I (a).** Various spectroscopic parameters of the dyes used.

| Dye | (a) $\tau \pm 0.01$ (ns) | (b) QY | (c) $J_{self} \pm 0.01$ ($\times 10^{13}$ M$^{-1}$cm$^3$) | (d) $R_{0self} \pm 0.1$ (Å) |
|---|---|---|---|---|
| D | 3.43 | 0.38[e] | 0.05 | 24.4 |
| $R_1$ | 2.85 | 0.29[f] | 1.00 | 38.4 |
| $R_2$ | 3.65 | 0.69[g] | 0.30 | 34.1 |
| $A_1$ | 1.18[h] | 0.03[i] | 30.00 | 47.3 |
| $A_2$ | 1.70[j] | 0.28[k] | 33.90 | 68.8 |

**Table I (b).** The spectral overlap and critical transfer distance for seven dye pairs

| Dye pair | Spectral overlap $J_{DA} \pm 0.1$ ($\times 10^{13}$ M$^{-1}$cm$^3$) | Critical transfer distance $R_{0A} \pm 0.1$ (Å) |
|---|---|---|
| CALN-HITCI (D-$A_2$) | 0.4 | 34.8 |
| CALN -DTTCI (D-$A_1$) | 0.5 | 35.8 |
| RhB-HITCI ($R_2$-$A_2$) | 3.2 | 50.7 |
| DODCI-HITCI ($R_1$-$A_2$) | 6.1 | 51.9 |
| CALN -DODCI (D-$R_1$) | 22.6 | 67.5 |
| DODCI-DTTCI ($R_1$-$A_1$) | 148.6 | 88.3 |
| CALN -RhB (D-$R_2$) | 148.6 | 92.4 |

(a) lifetime, (b) quantum yield, (c) intramolecular spectral overlap, and (d) intramolecular critical transfer distance of the dyes. (e) is taken from ref. [22], (f) from ref. [23], (g) from ref. [24], (h) and (i) from ref. [25], (j) from ref. [26], and (k) from ref. [27].

### 4.2.1 Donor- relay dye (D-$R_1$) system

The spectral overlap of this dye pair (D-$R_1$) is shown by the shaded region in the top panel of figure 2. The overlap integral and the corresponding critical energy transfer distance ($R_{0A}$) are given in Table I (b). The molar extinction coefficient of D is high (~5.5x10$^4$ M$^{-1}$cm$^{-1}$) and the ratio of the extinction coefficients of D to $R_1$ is 17 at the wavelength of excitation (488 nm). This clearly indicates the dominant absorption of the excitation light by the donor.

The concentration of D was kept constant as 0.1 mM, while that of $R_1$ was varied from 0.1 mM to 1 mM. It can be seen that on increasing the concentration of $R_1$, the fluorescence intensity of D decreases with a subsequent increase in the fluorescence intensity of $R_1$ (bottom panel, figure 2). Since there is no diffusion in the PVA matrix, this observation is an indication of thes diffusion-less, resonance energy transfer between D and $R_1$.

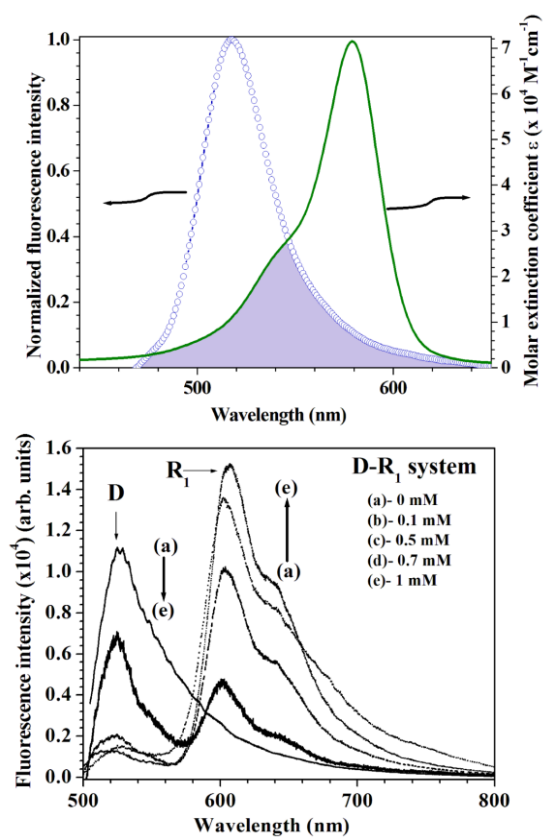

**Figure 2.** Fluorescence spectrum of donor (D) (○) and absorption spectrum of relay dye ($R_1$) (solid line) (top panel). The shaded region shows the spectral overlap. The fluorescence spectra of D-$R_1$ system are given in bottom panel for [D]= 0.1 mM. The relay dye concentrations are indicated in figure. The arrows indicate the peak positions for the fluorescence of D & $R_1$.

Quenching of fluorescence can also take place because of several other reasons such as electron transfer [28], collisional quenching by oxygen, iodine, acrylamide and a number of metal ions [29]. Hence, by using steady-state measurements alone it is not possible to know the exact mechanism. In addition, whenever there is a spectral overlap between donor fluorescence and acceptor absorption, the phenomenon of radiative energy transfer can take place [30]. Therefore, the time-resolved measurements help clarify the underlying mechanism and were carried out for these systems.

The fluorescence decay of D at the wavelength of 520 nm exhibits single exponential behaviour with a time constant of 3.43±0.01 ns. On increasing the concentration of $R_1$ the decay curves deviate from the single exponential nature and fit with the Förster type function (equation (3)). The results of the analysis of the decay profiles are summarised in table II. It can be seen that the critical transfer distance ($R_{0A}$) for $R_1$ concentration of 1 mM, is smaller (47 Å) than that calculated (67.5 Å) from the steady-state spectral overlap. The maximum efficiency obtained for this system was ~ 38 %, corresponding to an acceptor concentration of 1 mM.

### 4.2.2 Relay dye- acceptor ($R_1$-$A_1$) system

There exists considerable overlap between the fluorescence spectrum of $R_1$ and the absorption spectrum of $A_1$ leading to the large value of the overlap integral (Table I(b)). Therefore, this dye-pair of DODCI ($R_1$) and DTTCI ($A_1$) was selected as the relay dye-acceptor system. The overlap region is shown by the shaded region in the top panel, figure 3.

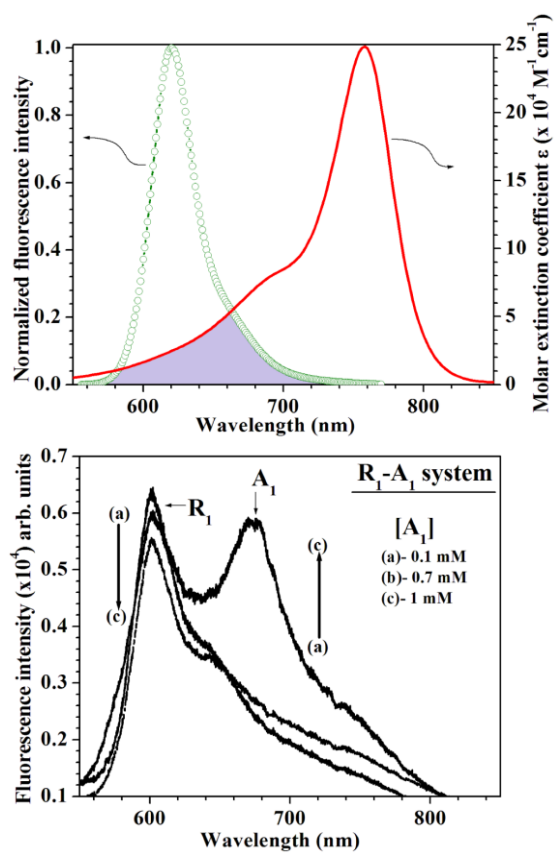

**Figure 3.** Fluorescence spectrum of relay dye ($R_1$) (○) and absorption spectrum of acceptor ($A_1$) (solid line) (top panel). The shaded region shows the spectral overlap. The fluorescence spectra of $R_1$-$A_1$ system are given in bottom panel for [$R_1$]= 0.1 mM. The acceptor concentrations are indicated in figure. The arrows indicate the peak positions for the fluorescence of $R_1$ & $A_1$.

Steady-state measurements have indicated the presence of FRET between $R_1$ and $A_1$ (bottom panel, figure 3). However, it was noticed that even in the presence of excitation energy transfer at higher acceptor concentration, the fluorescence intensity of the $A_1$ is not prominent. This is due to the low value of the quantum yield of NIR dyes. An increased probability of non-radiative decay as well as a decreased probability of radiative transitions at longer wavelength has been reported for most of the NIR dyes [31, 32]. Also it can be seen that the D-D overlap integral for the longest wavelength acceptors are more than the other dyes. This large self-absorption effects play an important role in deciding the final efficiency of FRET in the cascaded system.

**Table II.** The decay profile analysis of the D-$R_1$, $R_1$-$A_1$, and D-$A_1$ systems ($\lambda_{exc}$= 470 nm, $\lambda_{em}$= 520 nm).

| [Acceptor] (mM) | Single exponential fit | | Förster fit | | | $R_{0A}\pm0.1$ (Å) | $\eta$ (%) |
|---|---|---|---|---|---|---|---|
| | $\tau_D$ (ns) | $\chi^2$ | $\tau_D$ (ns) | $\gamma_{DA}$ | $\chi^2$ | | |
| **D-$R_1$ System** | | | | | | | |
| 0.5 | 3.10 | 1.10 | | | | | |
| 1 | 2.35 | 1.37 | 3.19 | 0.232 | 1.11 | 47 | 38.5 |
| **$R_1$-$A_1$ System** | | | | | | | |
| 0.1 | 1.81 | 1.15 | | | | | |
| 1 | 1.757 | 1.589 | 3.253 | 0.446 | 1.035 | 66 | 50.9 |
| **D-$A_1$ System** | | | | | | | |
| 1 | 3.83 | 1.12 | | | | | |
| 3 | 3.05 | 1.21 | 4.31 | 0.231 | 1.03 | 32.5 | 38.4 |

Table II gives the results of time-resolved measurements for $R_1$-$A_1$ system. The maximum efficiency is observed to be ~51% at [$A_1$] = 1mM. However, there is some difference in the $R_{0A}$ values obtained by time-resolved analysis is ~ 66 Å and spectroscopic measurements (~88 Å).

**4.2.3 Donor- acceptor (D-$A_1$) system**

The CALN (D) and DTTCI ($A_1$) dye system has low value of the overlap integral (overlap region is shown by the shaded region of the top panel of figure 4) and the critical transfer distance. This is due to the large shift between the donor fluorescence and acceptor absorption (panel (A), figure 1). The fluorescence spectra of D with varying concentrations of $A_1$ indicate that the fluorescence yield of the latter is small even at its highest concentration (bottom panel, figure 4).

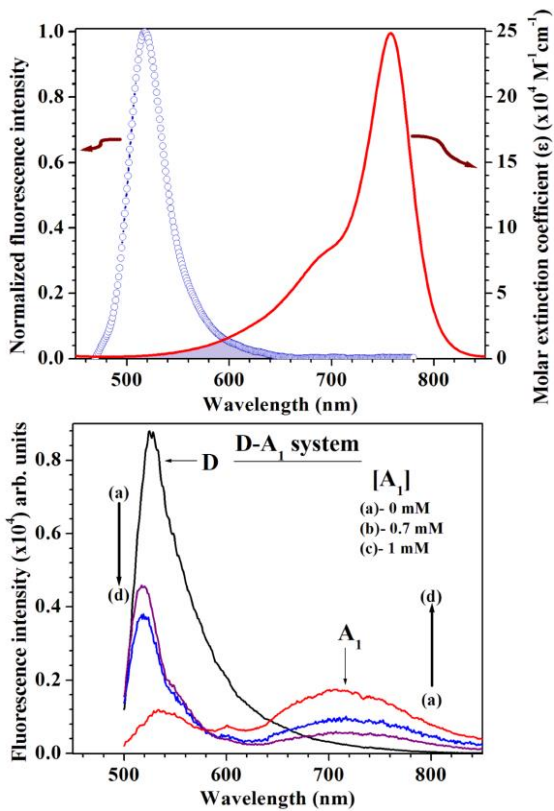

**Figure 4.** Fluorescence spectrum of donor (D) (○) and absorption spectrum of acceptor ($A_1$) (solid line) (top panel). The shaded region is the spectral overlap. The fluorescence spectra of D-$A_1$ system are given in bottom panel for [D]= 0.1 mM. The acceptor concentrations are indicated in figure. The arrows indicate the peak positions for the fluorescence of D & $A_1$.

The fluorescence decay curves were recorded for this system at donor emission wavelength of 520 nm. The fluorescence decay of D exhibits single exponential behaviour with a time constant of 3.43±0.01 ns (Table II). For lower concentrations of $A_1$ (up to ~3 mM), the decay curves follow the single-exponential behavior. A Förster function fit is obtained for higher concentrations. The $R_{0A}$ value obtained from the time-resolved data (~32.5 Å) is slightly less than that obtained from spectroscopic measurements (~35.8 Å). For a concentration of [$A_1$] = 3 mM, the FRET efficiency is estimated to be 38.4%.

### 4.3. Cascaded system of three dyes

The fluorescence spectra for CALN-DODCI-DTTCI (D-$R_1$-$A_1$) system by exciting at 488 nm is given in figure 5. For the cascaded system, the concentrations of D and $R_1$ were kept constant as 0.1 mM and 0.3 mM, respectively. It can be seen that, as the $A_1$ concentarion is increased from 0 to 1 mM, the fluorescence intensity of D decreases with a concomitant increase in the intensity of $A_1$. While fluorescence yield of $A_1$ was small for D-$A_1$ system, it is increased with the presence of the relay dye $R_1$.

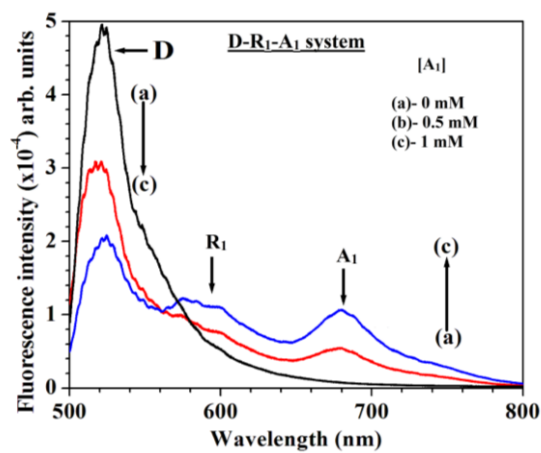

**Figure 5.** Fluorescence spectra of D-$R_1$-$A_1$ system for [D] = 0.1 mM, and [$R_1$] = 0.3 mM and acceptor concentrations are as indicated in the figure. The arrows indicate the peak positions for the fluorescence of D, $R_1$, & $A_1$.

The fluorescence decay curves of this system were also recorded at the donor emission wavelength (520 nm). As shown in Table III, the decay profiles are single exponential for lower concentrations of $A_1$. When the concentrations of $R_1$ and $A_1$ are increased, the decay profiles fit with Förster function (figure 6). The values of critical transfer distance ($R_{0A}$) and energy transfer efficiency ($\eta$) were calculated (equation (4) and (5)) from the value of reduced concentration ($\gamma_{DA}$) obtained by fitting the decay profiles. Corresponding to a typical acceptor concentration of 1 mM, the efficiency for this system was calculated to be ~42% (Table III) while there was no evidence of FRET in the case of D-$A_1$ system corresponding to the same acceptor concentration. Compared with the D-$A_1$ system with $A_1$ concentration of 3 mM, in the case of D-$R_1$-$A_1$ system with $A_1$ concentration of even 1 mM, the obtained efficiency is more. This indicates that the presence of the relay dye $R_1$ enhances the energy transfer from D to $A_1$.

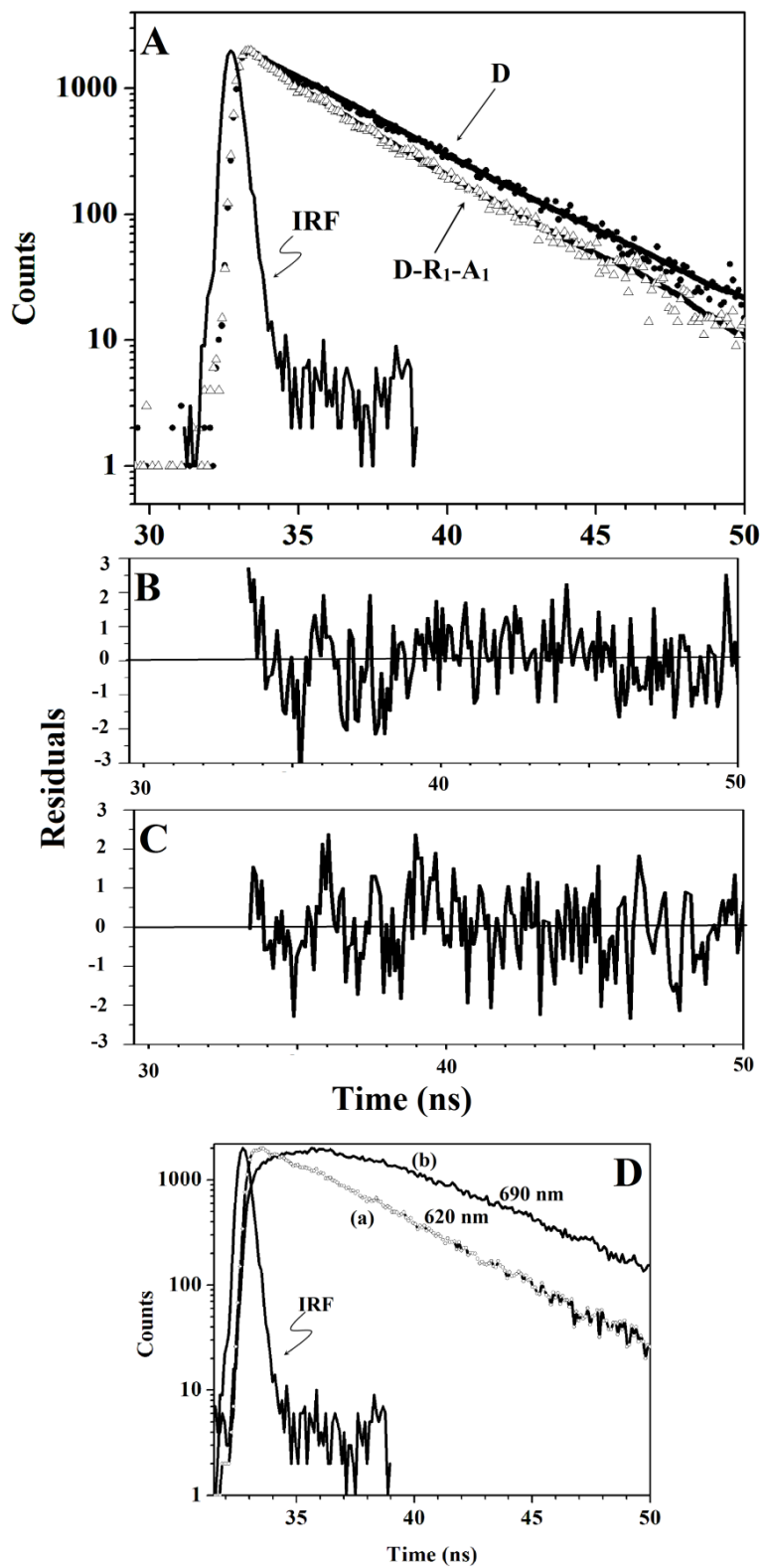

**Figure 6.** Panel A shows the fluorescence decay curves of D (0.1 mM) (●) and the D-$R_1$-$A_1$ system (Δ) at 520 nm with [$R_1$] = 1 mM and [$A_1$] = 5 mM. The solid lines show the single exponential fit for donor decay and Förster fit for the decay of the D-$R_1$-$A_1$ system, respectively. Panels B and C respectively show the residuals of the single exponential and Förster fits. Panel D shows fluorescence decay profiles of the D-$R_1$-$A_1$ system at the emission wavelength of 620 nm (a), and 690 nm (b) for the concentration same as in panel A.

**Table III**. The results of the time-resolved analysis with single exponential and Förster fit various dye systems. ($\lambda_{exc}$= 470 nm, $\lambda_{em}$= 520 nm)

| System | Concentrations (mM) | | Single exponential fit | | Förster fit | | | $R_{0A}\pm0.1$ (Å) | $\eta$ (%) |
|---|---|---|---|---|---|---|---|---|---|
| | | | $\tau_D$ (ns) | $\chi^2$ | $\tau_D$ (ns) | $\gamma_{DA}$ | $\chi^2$ | | |
| **D-R$_1$-A$_1$** | [D] = 0.1 [R$_1$]= 0.3 | [A$_1$]= 0 | 3.43 | 1.08 | | | | | |
| | | [A$_1$]= 1.0 | 2.78 | 1.24 | 3.79 | 0.325 | 1.00 | 66.3 | 41.3 |
| **D-R$_1$-A$_2$** | [D]= 0.1 [R$_1$]= 0.3 | [A$_2$]= 0 | 3.43 | 1.08 | | | | | |
| | | [A$_2$]= 2.0 | 1.15 | 3.41 | 1.68 | 2.511 | 1.03 | 96 | 92.9 |
| **D-R$_2$-A$_2$** | [D]= 0.1 [R$_2$]= 0.5 | [A$_2$]= 0 | 3.43 | 1.08 | | | | | |
| | | [A$_2$]= 2.0 | 2.40 | 1.83 | 3.33 | 0.639 | 1.02 | 24.5 | 62.4 |

The decay profile of D-R$_1$ system, ($\lambda_{em}$= 620 nm) with [D] = 0.1 mM & [R$_1$] = 1 mM fits with $\tau_1$= 0.93 ns, $\tau_2$= 2.97 ns with coefficients -0.31 and 0.35, respectively ($\chi^2$= 1.04).

The decay profiles recorded at 620 nm and 690 nm were found to have the contributions of acceptor rise time as shown in panel D of figure 6.

### 4.4. Comparison with other cascaded FRET systems

In order to see the effect of various spectral parameters such as spectral overlap, and QY on the cascaded FRET efficiency, similar study was carried out with two other three-dye systems: CALN-DODCI-HITCI (D-$R_1$-$A_2$) and CALN-RhB-HITCI (D-$R_2$-$A_2$). The results of the time-resolved studies of these systems and associated component systems are summarised in table III. The overall transfer efficiency is maximum in the case of the second system (D-$R_1$-$A_2$). It was noticed that, compared with the first two systems, the third one (D-$R_2$-$A_2$) is better in terms of the fluorescence yield. This is because of the higher values of QY of $R_2$ and $A_2$ compared with that of $R_1$ and $A_1$, respectively. So, the selection of dyes with higher QY will benefit in getting more yield in applications where light needs to be converted from a shorter wavelength region to a longer wavelength region.

A high value of the spectral overlap in any one cascaded component system alone is not a sufficient condition to get a high value of overall transfer efficiency. The QY of the relay dye and the acceptor plays an important role in the final yield at longer wavelength. Another important factor is the orientation factor. It is to be noted that the orientation of the transition dipoles depend upon the molecular size, shape and symmetry. The value of $\kappa^2$ can vary between 0 and 4 even though it is taken as ⅔ in the randomized donor and acceptor dipole cases [33].

### 5. Conclusions

The Förster resonance energy transfer between a dye pair with and without the presence of a relay dye has been studied. The study reveals that the presence of the relay dye ($R_1$) enhances the efficiency of energy transfer from the donor (D) to the acceptor ($A_1$). Experimentally it has been noticed that, a large value of the spectral overlap of one side - either donor to relay dye or relay dye to acceptor-alone is not sufficient for making the overall efficiency better. An additional parameter in increasing the final FRET yield can be an acceptor ($A_1$) with a higher QY. Therefore, this approach of cascaded energy transfer by selecting suitable dye-pairs with better efficiency can be used for efficient materials used in technological applications including solar devices.

**Acknowledgement**

We thank Department of Science and Technology (DST), New Delhi and Council of Scientific and Industrial Research (CSIR), New Delhi for financial support. Authors thank Dr. S. Mohan, for reading this manuscript.